\documentclass[aps,pra,floatfix,tightenlines,twocolumn,amsmath,amssymb,nofootinbib]{revtex4}
\pdfoutput=1

\usepackage{amsmath,bbm,graphicx}
\usepackage{subfigure}
\usepackage[letterpaper=true,pdftex,pdfpagelabels=true,bookmarks = true,colorlinks = true, citecolor=blue, urlcolor=cyan]{hyperref} 

\newcommand{\bra}[1]{\ensuremath{\left\langle #1\right\vert}}
\newcommand{\ket}[1]{\ensuremath{\left\vert #1\right\rangle}}

\newcommand{\hsp}[1]{\hspace{#1 em}}
\newcommand{\sqz}{\hsp{-0.1}}
\newcommand{\ketbra}[2]{\left\vert{#1}\right\rangle \sqz\sqz\sqz \left\langle{#2}\right\vert}

\def\>{\rangle}
\def\<{\langle}
\newcommand{\dg}{\ensuremath{^{\circ}}}

\def\iden{\mathbbmss{1}}
\newcommand{\PhiPlusPol}{\frac{1}{\sqrt{2}}\left(\ket{HH}+\ket{VV}\right)}

\def\rhoout{\ensuremath{\rho^{\text{out}}}}
\def\rhooutm{\ensuremath{\rho_{m}^{\text{out}}}}

\def\rhoexp{\ensuremath{\rho^{\text{exp}}}}
\def\target{\ensuremath{\rho^{\text{tar}}}}

\newcommand{\tr}{\text{Tr}}

\graphicspath{{Figures/}{rsp_paper/}}

\begin{document}
\title{Derivation and experimental test of fidelity benchmarks for remote preparation of arbitrary qubit states}
\date{\today}

\author{N. Killoran}
\affiliation{Institute for Quantum Computing and Department of Physics \&
Astronomy, University of Waterloo, Waterloo, Canada, N2L 3G1}
\author{D.N. Biggerstaff}
\affiliation{Institute for Quantum Computing and Department of Physics \&
Astronomy, University of Waterloo, Waterloo, Canada, N2L 3G1}
\author{R. Kaltenbaek}
\affiliation{Institute for Quantum Computing and Department of Physics \&
Astronomy, University of Waterloo, Waterloo, Canada, N2L 3G1}
\author{K.J. Resch}
\affiliation{Institute for Quantum Computing and Department of Physics \&
Astronomy, University of Waterloo, Waterloo, Canada, N2L 3G1}
\author{N. L\"{u}tkenhaus}
\affiliation{Institute for Quantum Computing and Department of Physics \&
Astronomy, University of Waterloo, Waterloo, Canada, N2L 3G1}

\begin{abstract}
Remote state preparation (RSP) is the act of preparing a quantum state at a remote location without actually transmitting the state itself. Using at most two classical bits and a single shared maximally entangled state, one can in theory remotely prepare any qubit state with certainty and with perfect fidelity. However, in any experimental implementation the average fidelity between the target and output states cannot be perfect. In order for an RSP experiment to demonstrate genuine quantum advantages, it must surpass the optimal threshold of a comparable classical protocol. Here we study the fidelity achievable by RSP protocols lacking shared entanglement, and determine the optimal value for the average fidelity in several different cases. We implement an experimental scheme for deterministic remote preparation of arbitrary photon polarization qubits, preparing 178 different pure and mixed qubit states with an average fidelity of 0.995. Our experimentally-achieved average fidelities surpass our derived classical thresholds whenever the classical protocol does not trivially allow for perfect RSP.
\end{abstract}

\maketitle

\section{Introduction}\label{sec:intro}
The field of quantum information processing has revealed many communication and computational protocols which can theoretically outperform their classical counterparts \cite{bouwmeester00a,nielsen00a}. Among the most famous is quantum teleportation \cite{bennett93a}, wherein Alice uses pre-shared entanglement and limited forward classical communication to produce an arbitrary unknown quantum state at Bob's location. Another example is \textit{remote state preparation} \cite{lo00b, pati00a} (RSP), a variant of teleportation where Alice has full knowledge of the state she intends to prepare at Bob's location. 
RSP protocols have several practical applications including forming part of deterministic arbitrary single-photon sources \cite{jeffrey04a} or efficient, high-fidelity quantum repeaters \cite{rosenfeld07a}.

However, due to the practical limitations of imperfect devices, no RSP experiment can yield remotely-prepared output states which \textit{exactly} match the intended states. Indeed, we should be satisfied when the output states have a high fidelity 
 with the intended states. This raises the question: how high must this fidelity be, on average, for an experiment to demonstrate a genuine quantum advantage? In other words, if we restrict Alice and Bob to a comparable, fixed amount of classical communication--but no shared entanglement--what is the optimal average RSP fidelity they could achieve? It is only when an experiment surpasses such a classical threshold that we can be 
sure of having demonstrated verifiable advantages to quantum communication. 

In several early publications on teleportation, thresholds are given to justify which results are genuinely in the non-classical regime \cite{furusawa98a,boschi98a,braunstein00a,braunstein01a}. For example, for the teleportation of qubit states, average fidelities higher than $\frac{2}{3}$ are not possible with only classical resources \cite{barnum95a}. To the best of our knowledge, such thresholds have neither been published nor tested for RSP. This paper then has two main objectives: First, we examine the limits on RSP with and without shared entanglement. Dependent on the target states and the allowed communication resources we derive several benchmarks separating genuinely quantum results from those which can be achieved with only classical communication. Second, we report and implement a new, fully-deterministic protocol for the remote preparation of arbitrary photon polarization states with high fidelity. Our protocol relies on generalized measurements (POVMs) and demonstrates several distinct advantages over previous experiments. In comparison with our derived benchmarks, our experimental data surpasses the limits of classical communication in all possible instances.

The remainder of this paper is organized as follows: In section \ref{sec:theory}, we outline the common framework for the RSP protocols examined in this work and flesh out the relevant theory in detail. We describe an entanglement-based protocol which theoretically achieves perfect fidelity between target and output qubit states using two classical bits of communication. We also analyze the optimal strategy in the ``classical'' case, where no entanglement is allowed. In section \ref{sec:thresholds} we evaluate the optimal classical thresholds and give benchmarks for several choices of pure target state ensembles, including finite, continuous, and mixed state ensembles. Section \ref{sec:experiment} describes our optical RSP experiment and compares our results to both pure and mixed state benchmarks. In section \ref{sec:conclusion} we conclude the paper.

\section{Theory}\label{sec:theory}
The goal of remote state preparation is to prepare a quantum state at a distant location, without sending the actual state. Alice, the sending party, knows exactly the target state $\target$ that she wants Bob, the receiving party, to have.
Several features are usually desired in an RSP protocol: Bob should need limited or zero knowledge of the state Alice is trying to prepare, and the required communication resources (classical and/or quantum) should be limited.  Perhaps most importantly, the protocol should yield output states $\rhoout$ at Bob's location which closely match the target states $\target$ which Alice intended to prepare.
There is no universally preferred measure for evaluating protocol performance, but in benchmarking situations where we want target and output states to match, the quantum fidelity \cite{jozsa94a} is a suitable choice, given by
\begin{equation}\
\label{eq:fidelity}
F(\sigma,\tau) = \left[\tr\left(\sqz \sqrt{\sqrt{\sigma}\hsp{0.2}\tau\sqrt{\sigma}}\hsp{0.2} \right)\right]^2.
\end{equation}
Ideally, the fidelity should be $F(\target,\rhoout)=1$ for any target state.

In order to make meaningful comparisons, we need a common framework to test the performance of RSP protocols and experiments. We imagine that Alice and Bob are challenged with the following task: Both parties are given full prior knowledge of some fixed ensemble of target states $\{ \rho_{\alpha}^{\text{tar}},p_{\alpha}\}$, and may coordinate beforehand on their strategy. To begin, Alice samples from the ensemble and, with probability $p_{\alpha_0}$, she picks the index $\alpha_0$. Unlike teleportation, Alice accesses the state \textit{index}, not the state, though she has complete 
information about the state and may prepare herself a copy if desired. She communicates a message to Bob, sending a limited number $c$ of classical bits (cbits). Bob then prepares an output state $\rho_{\alpha_{0}}^{\text{out}}$. Their goal is for the output states to match the target states with the highest possible quantum fidelity, on average, i.e. to maximize the quantity
\begin{equation}
 \langle F \rangle = \displaystyle\sum_\alpha p_\alpha F(\rho_{\alpha}^{\text{tar}},\rho_{\alpha}^{\text{out}}).
\end{equation}
We will be considering the situation where the target ensemble consists of a finite number of states as well as that where the target 
ensemble forms a continuum. In the latter situation, the above sum and probabilities are generalized to an integral and probability densities, respectively.

\begin{figure}
	\centering
		\includegraphics[width=1 \columnwidth]{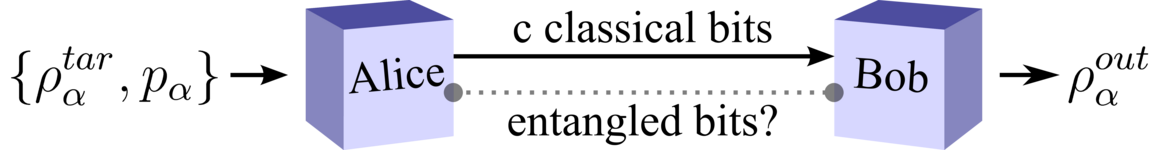}
	\caption{(Color online) Evaluating remote state preparation protocols. Alice samples a state $\rho^{\text{tar}}$ from a given distribution of target states and Bob aims to prepare a closely matching state. In classical RSP protocols, Alice may send only a limited number of classical bits to Bob. In quantum RSP protocols, the parties also share some pre-distributed entanglement. Their goal is to maximize $\<F(\target,\rhoout)\>$, the RSP fidelity averaged over the entire target distribution.
	}
	\label{fig:schematic}
\end{figure}

We are concerned in this work with two types of remote state preparation, which we call the ``quantum case'' and the ``classical case''. These labels refer to the communication resources allowed, and not the state prepared, which is always quantum mechanical. In the quantum case, Alice and Bob 
share a pre-distributed entangled state to help with their task. In the classical case, no initial quantum correlations between Alice and Bob are allowed. In both 
cases, once a target state has been selected, only $c$ cbits may be sent, and this classical communication is only permitted one way, from Alice to Bob. We will now investigate both of these cases separately.

\subsection{Quantum RSP}\label{ssec:quantum_rsp}
In this section we discuss the abilities and limitations of several quantum RSP protocols and briefly survey previous experimental implementations.
We confine the discussion to the remote preparation of qubit states, although some of the results generalize to higher dimensions. In all the protocols discussed, Alice must implement a measurement on her qubit from a shared entangled pair. We term a quantum RSP protocol \textit{deterministic} if the protocol succeeds for every outcome of this measurement. Furthermore we differentiate between those quantum RSP protocols where Alice can prepare any \textit{arbitrary} (pure or mixed) qubit at Bob's location, and those protocols which require that Bob have some foreknowledge of the state (e.g. that it be from some particular ensemble which forms a proper subset of all possible qubit states).
We then present the protocol employed in our experiment, which is deterministic and allows the preparation of arbitrary qubits.

An arbitrary qubit state $\rho$ can be expressed in terms of the $2\times2$ Pauli matrices:
\begin{equation}
 \rho = 
 \frac{\iden+\overrightarrow{r}\cdotp \overrightarrow{\sigma}}{2},
\end{equation}
where $\overrightarrow{\sigma}=(X,Y,Z)$ and $\overrightarrow{r}$ is the Bloch vector which uniquely identifies the state according to its position in the Bloch sphere. 
Alternately, the same qubit can be written as 
\begin{equation}
\rho(\phi,\theta,r) = r\ketbra{\psi}{\psi} + (1-r)(\iden/2),
\end{equation} 
where 
\begin{equation}\ket{ \psi ( \phi , \theta )} =\cos{(\phi/2 ) } \ket{0} +	e^{i \theta}  \sin{ ( \phi/2 ) } \ket{1}.
\end{equation}
Here $\phi$ and $\theta$ are the polar and azimuthal angles of $\rho$ in the Bloch sphere representation, respectively, and $r = |\overrightarrow{r}| = \sqrt{2\left(\tr\rho^2-\frac{1}{2}\right)}\in [0,1]$ is the radius of the state's Bloch vector.

Most RSP (and teleportation) protocols begin with the assumption that Alice and Bob share an initial supply of maximally-entangled qubit pairs,
usually one of the Bell states $\ket{\Phi^{\pm}}=\frac{1}{2}(\ket{00}\pm{\ket{11}})$ and $\ket{\Psi^{\pm}}=\frac{1}{2}(\ket{01}\pm{\ket{10}})$. Shared maximally-entangled pairs are sometimes called ebits.

Lo first proved that in the asymptotic (large-$N$) limit, Alice can deterministically prepare $N$ known pure qubit states from certain (restricted) ensembles at Bob's location using half the classical communication required for the teleportation protocol \cite{lo00b}; Pati \cite{pati00a} provided an explicit deterministic protocol whereby a single pure target qubit state $\target=\ketbra{\psi}{\psi}$ from such an ensemble can be remotely prepared with only one cbit and one ebit. The basic idea is as follows: Alice and Bob decide beforehand on an ensemble of states consisting of a single great circle on the Bloch sphere, specified by a Bloch vector $\hat{n}$. For each remotely prepared qubit, they share a singlet state $\ket{\Psi^-}$. Alice projects her entangled qubit into the basis $\{\ket{\psi},\ket{\psi^\perp}\}$. If the result is $\ket{\psi^\perp}$, Bob's qubit will be in state $\ket{\psi}$ as desired; if Alice's result is $\ket{\psi}$, Alice's transmitted cbit instructs Bob to perform the basis-specific {\sc not} operation on his qubit via rotating by $\pi$ about the $\hat{n}$ axis, thereby transforming his qubit $\ket{\psi^\perp}\mapsto\ket{\psi}$. However, if Alice wishes to remotely prepare an arbitrary pure qubit, not from a pre-specified great circle, Bob cannot reliably flip his qubit when he ends up with $\ket{\psi^\perp}$ due to the non-unitarity and thus non-physicality of a universal-\textsc{not} operation \cite{buzek99a}, and therefore the protocol is non-deterministic with only 50\% success probability.

Lo conjectured that two cbits transmitted from Alice to Bob would be necessary and sufficient for the deterministic remote preparation of an arbitrary qubit state with only one ebit \cite{lo00b}. This result was proven in Ref. \cite{leung03a} and, under more general conditions, in Ref. \cite{hayashi03a}. Many other papers have further investigated the trade-off between required cbits, ebits, and qubits for RSP (see e.g. Refs. \cite{bennett01b,hayden02a,abeyesinghe03a,berry03a,ye04a,berry04a,bennett05a}).

RSP protocols have been implemented to varying degrees in several experiments employing systems including nuclear magnetic spins \cite{peng03a}, coherent superpositions of photonic Fock states \cite{babichev04a}, atom-photon entanglement \cite{rosenfeld07a}, and polarization-entangled photon pairs \cite{jeffrey04a,ericsson05a,peters05a,xiang05a,liu07a,wu09a}. Among these, most of the employed protocols enabled the preparation of arbitrary pure states with 50\% success probability--or alternately, the deterministic preparation of qubits from specific, restricted ensembles. Several also enabled preparation of some mixed states \cite{jeffrey04a,ericsson05a,xiang05a}, and some allowed control of all three parameters $\{\phi,\theta,r\}$ required to prepare arbitrary pure or mixed states \cite{peters05a,liu07a}. Earlier, refs. \cite{rosenfeld07a,liu07a} and, while we were preparing this manuscript, \cite{wu09a} successfully implemented a generalized measurement on Alice's qubit which should allow Bob to perform a unitary correction and achieve the desired target state regardless of Alice's measurement outcome. However, in none of these papers is the required unitary actually implemented. To the best of our knowledge we present the first experimental implementation of a \textit{fully deterministic} RSP protocol enabling the preparation of arbitrary (mixed and pure) qubit states.

Note however that in our actual experiment, Bob does not necessarily register a detection event every time that Alice detects a photon. This is due to coupling losses and detector inefficiency, which are unrelated to the efficiency of the RSP protocol itself. These experimental considerations necessitate postselection on coincident detection events between \textit{any} of Alice's four measurement outcomes and Bob's detector. Only if a coincidence occured can one infer that Alice and Bob shared an entangled pair (ebit), a prerequisite for quantum RSP. This differs from the protocol employed in e.g. Ref \cite{peters05a}, which employs postselection to detect the ebit \emph{and} for a specific measurement outcome. In our experiment the postselection is only used to verify a shared ebit, and the protocol then functions deterministically, succeeding for all of Alice's measurement outcomes \cite{prevedel07a}.

Our protocol makes use of the Bell state $\ket{\Phi^+}$ and (at most) two cbits to remotely prepare an arbitrary state $\rho (\phi,\theta,r)$. For any pure state $\ket{\psi}$, the Bell state $\ket{\Phi^+}_{AB}$ can be written as $\frac{1}{2}\sum_{m=1}^4 \sigma_m^A \sigma_m^B \ket{\psi^*_A} \ket{\psi_B}$,
where $\ket{\psi^*(\phi,\theta)}$ is the complex conjugate of $\ket{\psi(\phi,\theta)}$ in the computational basis, and $\sigma_m^{A(B)} \in \{\iden,X,XZ,Z\}$ are Pauli operators acting on Alice's (Bob's) qubit.
 
First, consider the case where Alice would like to help Bob remotely prepare a pure state $\rho(\phi,\theta)=\ketbra{\psi}{\psi}$.  She first performs a generalized measurement on her qubit, specifically a positive operator-valued measure or POVM \cite{nielsen00a,barnett97a,ahnert05a}, defined by the elements 
\begin{equation}
\label{eq:POVM}
\{E_m(\phi,\theta)\} = \frac{1}{2}\{\sigma_m\ketbra{\psi^*}{\psi^*}\sigma_m^\dagger\}.
\end{equation}
Dependent on the outcome $m \in \{0,...,3\}$ obtained, Bob's qubit
will be left in the state $\sigma_m\rho\sigma_m^\dagger$. Alice then encodes
the outcome $m$ in two cbits and transmits the resulting message to Bob. By implementing $\sigma_m$ on his qubit, Bob will deterministically recover $\rho$.

The generalization of this scheme for preparing arbitrary mixed states is
quite straightforward. If Alice sends the same message to Bob regardless of her measurement outcome, his qubit will be left in the maximally mixed state $\frac{1}{2}\iden$. In order to remotely prepare an arbitrary state $\rho (\phi,\theta,r)$, Alice performs the same POVM $\{E_m(\phi,\theta)\}$ as she would to prepare the pure state $\ketbra{\psi(\phi,\theta)}{\psi(\phi,\theta)}$. However, she only transmits the correct message encoding the POVM outcome to Bob with probability $r$. Otherwise she sends a particular message, regardless of the outcome obtained. Thus with probability $r$ Bob's qubit ends up in $\ketbra{\psi}{\psi}$, and with probability $(1-r)$ he has $\frac{1}{2}\iden$, as desired. Due to the unequal distribution of probabilities among the messages, the classical communication cost required to prepare mixed states, as measured by the Shannon entropy \cite{shannon49a}, will thus be less than for pure states.
This cost will range from 0 cbits for preparation of the maximally mixed state to 2 cbits for pure states.\footnote[1]{In particular the Shannon entropy of the communication required for preparing a state of purity $r$ will be $H(r)=2-\log_2(4-3r)+\frac{3r}{4}\log_2\left(\frac{4-3r}{r}\right)$.}

\subsection{Classical RSP}\label{ssec:classical_rsp}
We now examine the classical case, where Alice and Bob share no entanglement. As our goal is to find the optimal achievable fidelity, we assume in this scenario that Alice and Bob are unencumbered by the imperfections of real-world devices. This assumption is in the spirit of security proofs for quantum key distribution, where any adversary Eve is assumed to be limited only by the laws of physics. It is only by surpassing the limits of this ideal scenario that an experiment can provably demonstrate genuine quantum advantages. Therefore, the one-way classical channel between Alice and Bob is assumed to be perfect, as is Bob's ability to prepare any desired output state. 

Although the experiment detailed in Section \ref{sec:experiment} is for qubit states, some of the results in this section hold equally well for states in any finite dimensional Hilbert space. We begin with no assumptions about the dimension except that it is finite, and we will specialize to qubits (dimension 2) when appropriate. Furthermore, we are primarily interested in the case where the target states are pure, $\rho_\alpha^{\text{tar}} = \ketbra{\psi_\alpha^{\text{tar}}}{\psi_\alpha^{\text{tar}}}$, so that the quantum fidelity is equal to the matrix element
\begin{equation}
 F(\rho_\alpha^{\text{tar}}, \rho_\alpha^{\text{out}}) = \bra{\psi_\alpha^{\text{tar}}}\rho_\alpha^{\text{out}}\ket{\psi_\alpha^{\text{tar}}}.
\end{equation}
Accordingly, we assume that the target ensemble consists of pure states $\{ \ket{\psi_\alpha^{\text{tar}}},p_\alpha \}$. In section \ref{sec:thresholds} we give benchmarks based on specific choices for this target ensemble. 

We now examine the question: what is the optimal RSP strategy when the parties share no quantum correlations, and Alice may only send $c$ cbits to Bob? For every target state $\ket{\psi_\alpha^{\text{tar}}}$, Alice sends a string of $c$ classical bits. We can label all messages of this type by a natural number $m(\alpha)=k\in \{0,1,...,2^{c}-1 \}$. In general, the message assignment may be either deterministic (e.g. $m(\alpha) = 3$) or probabilistic, i.e. $m(\alpha) = k$ with probability $q_k(\alpha)$, where for each $\alpha$, $\sum_k q_k(\alpha) = 1$. The probabilistic framework contains all deterministic strategies as special cases. Note that here we use `deterministic' to refer to Alice's messaging strategy whereas elsewhere it is used to refer to the success probability of the protocol (Sec. \ref{ssec:quantum_rsp}); in general our meaning will be clear from the context.

Upon receiving the message $k$, Bob prepares some output state $\rho^{\text{out}}_k$. A probabilistic messaging strategy would necessarily lead Bob to prepare a \textit{mixed} output state $\rho_{m(\alpha)}^{\text{out}} = \sum_k q_k(\alpha)\rho^{\text{out}}_k$ whenever state $\ket{\psi_\alpha^{\text{tar}}}$ is chosen. Similarly, for a given message $k$, Bob may change the output state probabilistically. This strategy is naturally incorporated into our framework, where we allow the output states $\rho_k^{\text{out}}$ to be mixed.

To determine which choice of output states optimize the average fidelity, we rewrite it in terms of the $2^c$ unique messages:
\begin{align}
 \langle F \rangle 	= & \displaystyle\sum_{\alpha} p_\alpha\bra{\psi_\alpha^{\text{tar}}}\rho^{\text{out}}_{m(\alpha)}\ket{\psi_\alpha^{\text{tar}}}\notag\\
			= & \displaystyle\sum_{k=0}^{2^c-1}\displaystyle\sum_{\alpha} p_\alpha q_k(\alpha) \tr(\ketbra{\psi_\alpha^{\text{tar}}}{\psi_\alpha^{\text{tar}}} \rho_k^{\text{out}})\notag\\
			= & \displaystyle\sum_{k=0}^{2^c-1} p_k \tr(\overline{\rho_k} \rho_k^{\text{out}})
\end{align}
where $p_k = \sum_{\alpha} p_\alpha q_k(\alpha)$ is the probability of Alice sending message $k$ and $\overline{\rho_k} = \frac{1}{p_k}\sum_{\alpha} p_\alpha q_k(\alpha) \ketbra{\psi_\alpha^{\text{tar}}}{\psi_\alpha^{\text{tar}}}$ is a weighted average of the states where message $k$ might be sent. When the fidelity is written in this form, two notable features become apparent:

\begin{enumerate}
	 \item For each $k$, the quantity $\tr(\overline{\rho_k} \rho_k^{\text{out}})$ is upper bounded by the largest eigenvalue $\lambda^{\text{max}}_k$ of the average state $\overline{\rho_k}$; this can be achieved if Bob outputs the corresponding eigenstate $\rho_k^{\text{out}} = \ketbra{\lambda_{k}^{\text{max}}}{\lambda_{k}^{\text{max}}}$. Thus, the optimal output states give
	\begin{equation}\label{Fmax}
 		\langle F \rangle^{\text{max}} = \sum_{k=0}^{2^c-1} p_k \lambda_k^{\text{max}}.
	\end{equation}

	\item Since the optimal output states are pure (by point 1), the optimal messaging strategy must, therefore, be deterministic, not probabilistic. In other words, a unique message is sent for each target state. This corresponds to only one $q_k(\alpha)$ being non-zero for each $\alpha$. 
\end{enumerate}

Taking these two points into account greatly simplifies the structure of the fidelity optimization. Because the optimal message assignment is deterministic, the target ensemble is effectively split into $2^c$ disjoint partitions, depending only on the message $k\in \{ 0,1,...,2^{c-1} \}$. For each partitioning of the target ensemble, we can also calculate the optimal output state and the resulting fidelity value using Eq. (\ref{Fmax}). All that remains is to determine \textit{which partitioning} maximizes the value of Eq. (\ref{Fmax}). To clarify notation, we will henceforth use $k$ to label both a message and the partition of the target ensemble consisting of states for which that message is sent. Again, the meaning will be clear from the context.

In principle, for a finite number $n$ of target states, the remaining optimization problem only requires checking the value of Eq. (\ref{Fmax}) for each of the finite number of possible partitionings, which can be done by computer. However the number of possible partitionings scales exponentially in $n$, rendering this calculation unreasonable for more than about $n=10$ states. In section \ref{sec:thresholds} we outline an algorithm which efficiently provides bounds to Eq. (\ref{Fmax}).

\emph{Qubits.} If the states in question are qubits, we can put Eq. (\ref{Fmax}) into a simple geometric form. When expressed in its eigenbasis, a qubit state takes the form
\begin{equation}
 \rho = 
 \frac{1}{2}\begin{bmatrix}
  1+r & 0\\
  0 & 1-r
 \end{bmatrix}
\end{equation}
where $r$ is the radius of the state's Bloch vector. The largest eigenvector of a qubit is directly related to the radius: $\lambda^{\text{max}} = \frac{1+r}{2}$. 
For any deterministic partitioning of the target ensemble, we denote the average Bloch vectors by $\overrightarrow{r_k} = 
\frac{1}{p_k}\sum_{\alpha \in k}p_\alpha \overrightarrow{r_\alpha}$
 and their magnitudes by $r_k$. We find that the maximal average fidelity for qubits is given by

\begin{equation}\label{QubitFmax}
 \langle F \rangle^{\text{max}} = \frac{1}{2}\left( 1+\displaystyle\sum_{k=0}^{2^c-1}p_k r_k \right).
\end{equation}

Hence, for a given deterministic partitioning, the best average fidelity is determined by two sets of quantities: the probabilities $p_k$ of sending each message and the length of the average Bloch vectors $\overrightarrow{r_k}$ within each of the $2^c$ partitions.

In section \ref{sec:thresholds}, we will outline how to determine which choice of messages, i.e. which partitioning of the target ensemble, maximizes Eq. (\ref{QubitFmax}).

\begin{figure}
	\centering
		\includegraphics[width=1 \columnwidth]{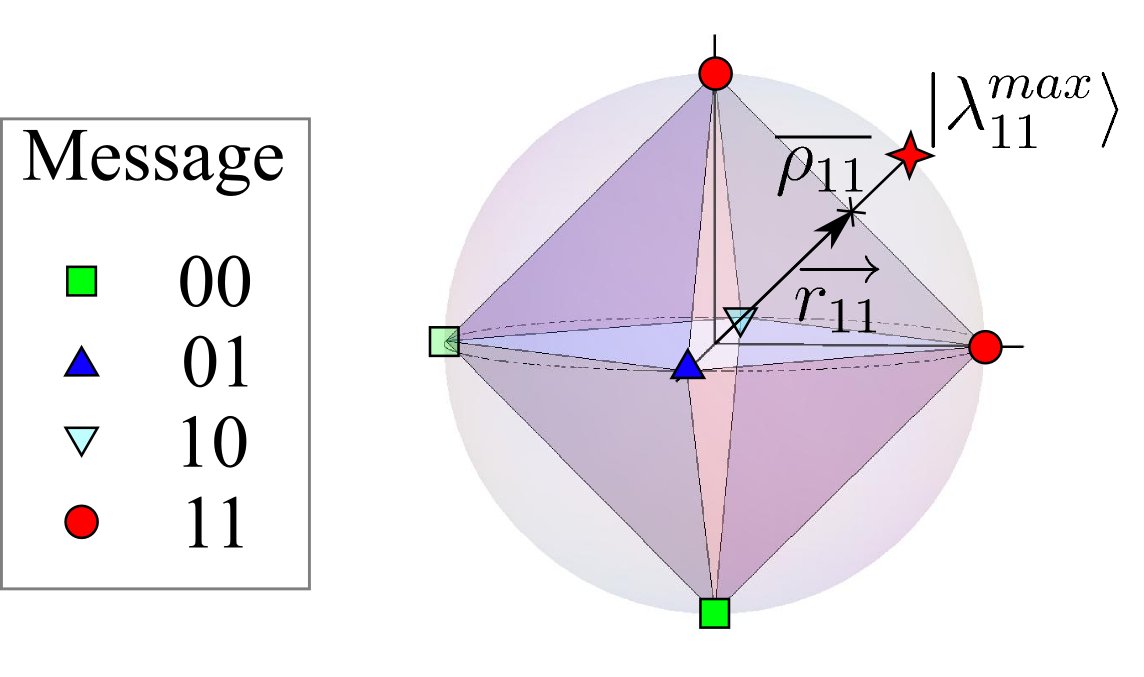}
	\caption{(Color online) Example of a possible classical remote state preparation strategy. The target ensemble consists of the 6 pure states $\{\ket{0}, \ket{1}, \frac{\ket{0}\pm\ket{1}}{\sqrt{2}},\frac{\ket{0}\pm i\ket{1}}{\sqrt{2}}\}$ (represented here as the vertices of an octahedron inscribed within the Bloch sphere) with equal probabilities. A possible partitioning strategy is given for the case where two cbits of classical communication are allowed, and the optimal output state for partition 11 is detailed.}
	\label{fig:partitioning}
\end{figure}

\section{Threshold calculations}\label{sec:thresholds}

\subsection{Finite ensembles}\label{ssec:finite}
Assume now that we have fixed a \textit{finite} ensemble of target states $\{ \rho_\alpha^{\text{tar}}, p_\alpha \}_{\alpha=1}^{n}$. It is clear that whenever $n\leq2^c$, the optimal classical protocol can achieve perfect fidelity since there is sufficient capacity in the message to uniquely label the state. The interesting cases have $n>2^c$. Given the results of the previous section, the optimum average fidelity can be determined by checking the value of Eq. (\ref{Fmax}) for all partitionings of the $n$ target states into $2^c$ disjoint subsets, but this can be inefficient even for modest values of $n$ and $c$. Alternatively, we search for an upper bound on the threshold which is easier to calculate. If an experiment surpasses the upper bound, it has surpassed the actual threshold.

We will now outline an efficient algorithm for determining such upper bounds. 
For this algorithm, we make the additional assumption that each target state has equal probability to be chosen from the target ensemble. 
We note that each partition contains some number $s$ of states and contributes one term to the sum in Eq. (\ref{Fmax}). Two different partitions with the same number of states may contribute differently to the average fidelity, depending on the arrangement of the states. However, for each number $s\in \{0,1,...,2^c-1\}$, there is a set of $s$ states which yields the maximal possible contribution $\langle F\rangle^{\text{max}}_s$. By using these maximal values in Eq. (\ref{Fmax}) instead of the actual values, we obtain an upper bound on the threshold.

The first step in the algorithm involves checking all partitions of size $s$ to find the maximal contribution $\langle F\rangle^{\text{max}}_s$. Next, we list all the ways in which $n$ elements can be divided into $2^c$ subsets. The order of the subsets does not matter, so for simplicity we can create our list in order of decreasing partition size. This list forms a table with $2^c$ columns. For each row $i$, we have a list of numbers $\{s_{ij}\}_{j=0}^{2^c-1}$ which sum to $n$.
To determine the upper bound, we calculate the quantity
\begin{equation}
 \langle F \rangle_i = \sum_{j=0}^{2^c-1}\frac{s_{ij}}{n}\langle F\rangle_{s_{ij}}^{\text{max}}.
\end{equation}
The highest $\langle F\rangle_i$ provides us with an upper bound on the optimal average fidelity. 

It may even be the case that the threshold is equal to the upper bound found via the above algorithm, especially if the target ensemble exhibits a high degree of symmetry. To verify this, one would have to find a specific partitioning which leads to the same value as the upper bound. On the other hand, if we can show through other arguments that the highest $\langle F\rangle_i$ is unachievable, then the second highest $\langle F\rangle_i$ provides a new, smaller upper bound. We will make use of both of these points below. 

Before proceeding, we pause to discuss the tradeoff between classical and quantum communication resources. The remote state preparation scheme outlined in sections \ref{ssec:quantum_rsp} and \ref{sec:experiment} uses one entangled qubit (ebit) and two cbits sent from Alice to Bob to remotely prepare pure qubit states and less than two cbits for mixed states. A classical analog might limit Alice to sending two cbits to Bob each run. However, it may be argued that to distribute the entangled qubit between Alice and Bob requires at least one use of a quantum channel. A more fair comparison scenario might then allow Alice one use of this quantum channel per run, but only to send classical information. In this scenario, Alice sends three cbits in total. Arguably, this is unnecessary as one could consider the entanglement to be distributed by Bob or by a third party. However, the more cbits Alice is allowed to transmit, the higher the average fidelity the parties can achieve, rendering the benchmark that much harder to surpass in experiment. For completeness and comparison purposes we henceforth consider both the two and three cbit cases.

\emph{Benchmarks.} Thus far no specific target ensemble has been chosen. We now examine several specific ensembles for comparison with experiment. We restrict ourselves to ensembles of pure qubit states with a uniform distribution: $p_i=\frac{1}{n}$. If our goal is to find benchmarks which are low enough to be experimentally surpassed, we should make the classical task as difficult as possible. Given the results above, this is accomplished by choosing ensembles of states which are maximally ``spread apart'', so that the average Bloch vector within any partition is as small as possible. 

An effective choice is to use the vertices of the \textit{Platonic solids} inscribed in the Bloch sphere as the target states. The Platonic solids are the tetrahedron, octahedron, cube, icosahedron, and dodecahedron, with 4, 6, 8, 12, and 20 vertices, respectively. Note that the orientation of these vertices with respect to a Cartesian reference frame does not matter in the classical case, but a specific choice must be made in an experiment. Also note that the tetrahedron states do not provide a surpassable benchmark for $c\geq 2$ because they can be prepared with perfect fidelity simply by assigning a unique message to each of the 4 states. Similarly, for three cbits, the benchmarks yielded by the tetrahedron, octahedron and cube ensembles are all trivially unity. For the other cases, however, we expect fidelity thresholds less than unity. 

\begin{figure*}
	\centering
		\includegraphics[width=2 \columnwidth]{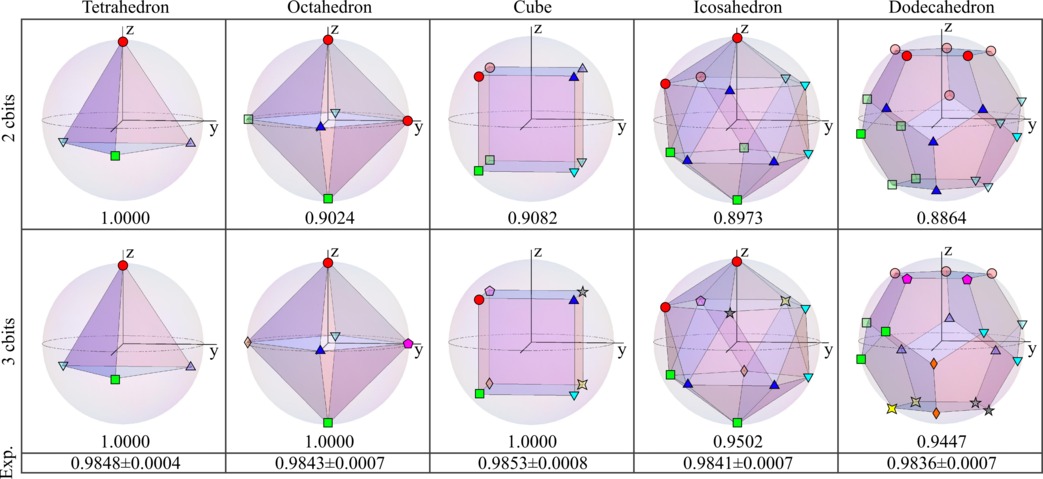}
	\caption{(Color online) Five examples of pure state target ensembles, given by the vertices of the Platonic solids inscribed within the Bloch sphere, with uniform probability distributions. For both two and three cbits message capacity, optimal partitioning strategies are shown, along with the corresponding optimal average fidelity benchmarks. States labeled with the same symbol (e.g. red circles) are in the same partition. The tetrahedron (two or three cbits), octahedron (three cbits), and cube (three cbits) examples can in principle be remotely prepared with perfect fidelity using only classical communication, whereas the remaining ensembles cannot. Experimentally achieved mean fidelities for these ensembles are given in the bottom row; the reported uncertainty is the standard error of the mean. For all non-unity benchmarks, the experimental values surpass the benchmarks for two (three) transmitted cbits by at least 96 (46) times the standard error of the mean.}
	\label{fig:optimal}
\end{figure*}

Indeed, using the algorithm above, we can calculate upper bounds on the remaining thresholds, all of which are less than unity. In fact, for every example studied except for one, the upper bounds were actually equal to the optimal classical thresholds. This was verified by finding explicit partitions such that Eq. (\ref{Fmax}) saturated the upper bounds. The one exception to this statement is the dodecahedron ensemble when $c=2$. In this case, the upper bound returned by the algorithm would only be possible if we could partition the dodecahedron vertices into four disjoint pentagons. This is geometrically impossible, so we can omit this upper bound. The next highest bound, consisting of partitions of size 6, 5, 5, and 4, is indeed possible. The optimal thresholds and their corresponding partitions, along with experimental results, are given in Fig. \ref{fig:optimal}.

\subsection{Continuous ensemble}\label{ssec:continuum}

Perhaps the most meaningful target ensemble is the uniform ensemble of all pure qubit states, i.e. the Bloch sphere itself. For the related problem of teleportation, the optimal classical strategy leads to an average fidelity of $\frac{2}{3}$ \cite{barnum95a}. Remote state preparation should be easier than teleportation, since Alice has complete knowledge of the state. The fidelity threshold should therefore be higher, though the threshold will also depend on how many communicated cbits are allowed. Therefore, demonstrating genuine non-classical behaviour experimentally is more difficult for RSP than for teleportation. 

We will now derive upper and lower bounds on the classical threshold for both two and three cbits. Many of the results for pure states from the previous section, suitably generalized, still hold here. Partitions will be denoted by $\Omega_k$, with their union forming the surface of the Bloch sphere, $\cup_k\Omega_k = S^2$.
The optimal average fidelity is still given by Eq. (\ref{QubitFmax}), but we make the modifications
\begin{align}
& p_k \rightarrow  \displaystyle\frac{1}{4\pi}\displaystyle\int_{\Omega_k} d\Omega = \frac{A_k}{4\pi},\notag\\
& \overrightarrow{r_k} \rightarrow  \displaystyle\frac{1}{p_k}\displaystyle\frac{1}{4\pi}\displaystyle\int_{\Omega_k}\overrightarrow{r_\alpha}d\Omega = \frac{1}{A_k}\int_{\Omega_k} \overrightarrow{r_\alpha}d\Omega,\label{integralfidelity}
\end{align}
where $A_k$ is the surface area of partition $k$.

To obtain lower bounds on the threshold, we simply choose a particular partitioning. For two cbits, we imagine that a tetrahedron is inscribed in the Bloch sphere and connect the four vertices by segments of great circles (note: this is not to be confused with use of platonic solids in previous section). This leads to four disjoint regions on the surface of the Bloch sphere which form our partitions. To calculate the optimal average fidelity for this arrangement, we integrate Eq. (\ref{integralfidelity}) and make use of the following equation for great circles in spherical coordinates: $\cot(\phi)=a\sin(\theta+c)$ (\cite{morgan05a}, Lemma 28.1). Here, $\phi\in[0,\pi]$ and $\theta\in[0,2\pi]$ are the polar and azimuthal angles, respectively, and $a$ and $c$ are constants determined by substituting two points which the great circle passes through. Using this relation, the bound can be worked out to be 0.8724. For three cbits, we use the eight octants as our partitions (equivalently, we connect the vertices of an inscribed octahedron). This straightforwardly gives a lower bound on the threshold of 0.9330. We conjuecture that these two lower bounds are the optimal values, but we cannot prove at this time.

To obtain upper bounds, we use an idea similar to the algorithm detailed in section \ref{ssec:finite}. If we can determine the maximal weighted average fidelity $\langle F \rangle^{\text{max}}_A$ achievable for a given surface area $A$ of the sphere, then we can calculate an upper bound using these values: 
\begin{equation}
 \langle F \rangle^{\text{max}} \leq \sum_{k=0}^{2^c-1}\langle F \rangle^{\text{max}}_{A_k} ~ s.t. ~ \sum_{k=0}^{2^c-1} A_k = 4\pi.
\end{equation}
 
Out of all possible configurations of a given partition with area $A_k$, a circular cap on the Bloch sphere gives the longest average Bloch vector, and hence the largest average fidelity. Also, the optimal distribution occurs when all partition areas are equal, $A_k = \frac{4\pi}{2^c}~\forall~k$ (see Appendix \ref{app:continuumbound} for proofs of these statements). Of course, it is only possible to cover the Bloch sphere with $2^c$ disjoint circular caps when $c=0\mbox{ or }1$, so the upper bounds for $c\geq 2$ are not achievable. 

Using the equations derived in Appendix \ref{app:continuumbound}, the upper bounds for $c=2$ and $c=3$ work out to be 0.8750 and 0.9375, respectively. Even these simple ideas yield tight bounds on the continuum thresholds for two and three cbits. To summarize: 
\begin{align*}
 0.8724 &\leq \langle F \rangle^{\text{max}} < 0.8750 \text{ for } c=2,\\
 0.9330 &\leq \langle F \rangle^{\text{max}} < 0.9375 \text{ for } c=3. 
\end{align*}

These numbers are significantly higher than the optimal classical teleportation fidelity of $\frac{2}{3}$. This confirms that, when restricted to classical communication only, the remote preparation of a known quantum state is indeed easier than the teleportation of an unknown quantum state. It is thus more difficult to demonstrate a genuine quantum advantage in an RSP experiment than in a teleportation experiment.

\subsection{Mixed states}\label{ssec:mixedstates}

Here we consider the same type of qubit ensembles as in section \ref{ssec:finite}, but with the modification that every state in an ensemble is a mixed state with Bloch vector length $r$. Unfortunately, if the target states are mixed states, finding classical thresholds is more complicated than in the pure state case. For instance, the optimal strategy is not necessarily one with deterministic messaging. Consider a target ensemble consisting of the three qubit states $\rho_{\alpha_1}^{\text{tar}} = \ket{0}\bra{0},~\rho_{\alpha_2}^{\text{tar}} = \ket{1}\bra{1},~\rho_{\alpha_3}^{\text{tar}} = \frac{1}{2}\iden$ with equal probability $\frac{1}{3}$. Alice sends messages according to the distribution 
\begin{align*}
   q_0(\alpha_1) = 1, ~ & q_1(\alpha_1) = 0\\
   q_0(\alpha_2) = 0, ~ & q_1(\alpha_2) = 1\\
   q_0(\alpha_3) = \displaystyle\frac{1}{2}, ~ & q_1(\alpha_3) = \displaystyle\frac{1}{2},
\end{align*}
and Bob prepares the two output states $\rho_0^{\text{out}} = \ket{0}\bra{0},~ \rho_1^{\text{out}} = \ket{1}\bra{1}.$ It is easy to see that this probabilistic messaging strategy, which uses only one cbit, allows Alice and Bob to remotely prepare any three of these states with arbitrarily high fidelity. 

In fact, for any target ensemble which is contained in the convex hull of $N \leq 2^c$ suitably chosen points, Alice and Bob can achieve an arbitrarily high fidelity by using a probabilistic messaging strategy. For instance, if Alice has access to two cbits, she could specify four pure states which form the vertices of a tetrahedron and prepare any state within this tetrahedron with perfect fidelity. Similarly, with three cbits, she could perfectly prepare any state located within a cube whose vertices were pure states. For example, consider a uniform dodecahedron ensemble with each state having Bloch radius $r$. For two (three) cbits, if this radius is not larger than the radius of a sphere inscribed in the tetrahedron (cube), then the ensemble can be prepared with perfect fidelity. For two (three) cbits, the insphere radius is $\frac{1}{3}$ ($\sqrt{\frac{1}{3}}$). Similar statements can be made for any ensemble with states of constant radius.

The possibility that the optimal strategy could involve probabilistic messaging renders the optimization trickier, as we can no longer use a partitioning argument to find the optimal value. Another approach is to focus on finding the optimal strategy which involves only deterministic messages. This is the special case where, for each target state $\rho_\alpha^{\text{tar}}$, only one of the $q_k(\alpha)$ is non-zero. The optimal value in this case, found by optimizing over output states, provides a lower bound to the true optimum. Unfortunately, this restriction does not fairly match with our experiment, where messages are probabilistically determined by measurement outcomes. However, surpassing this bound is at least a \emph{necessary} condition, if not a sufficient one, for any remote state preparation experiment to demonstrate non-classical advantages. 

Under this deterministic messaging assumption, the optimal choices of $\rho_k^{\text{out}}$ (see proof in Appendix \ref{app:mixedstates}) achieve a maximal average fidelity of
\begin{equation}\label{Fmax_mixed}
 \langle F \rangle^{\text{max}}_{determ.} = \frac{1}{2}\left( 1 + \sum_{k=0}^{2^c-1} p_k\sqrt{r_k^2+1-r^2} \right),
\end{equation}
where $p_k$ and $r_k$ are the same quantities as defined for pure qubit states. In general, this optimal value is achieved using mixed output states. The fidelity in Eq. (\ref{Fmax_mixed}) is modified from the pure state case, Eq. (\ref{QubitFmax}), by the additional term $1-r^2$ under the square root. Since this term is fixed beforehand, it does not change which partitioning of the target ensemble is optimal. In other words, whichever partitioning maximizes Eq. (\ref{QubitFmax}) for an ensemble of pure states will also maximize Eq. (\ref{Fmax_mixed}), the fidelity bound for the corresponding ensemble with Bloch radius $r$. Experimental data is compared with these theoretical bounds in Fig. \ref{fig:mixed}.

\section{Experiment}\label{sec:experiment}
We experimentally implemented the RSP protocol described in section \ref{ssec:quantum_rsp} using optical photons as our qubits. The computational basis states are $\ket{H}=\ket{0}$ and $\ket{V}=\ket{1}$, which respectively indicate indicate horizontal and vertical polarization.

\begin{figure}
	\centering
		\includegraphics[width=1 \columnwidth]{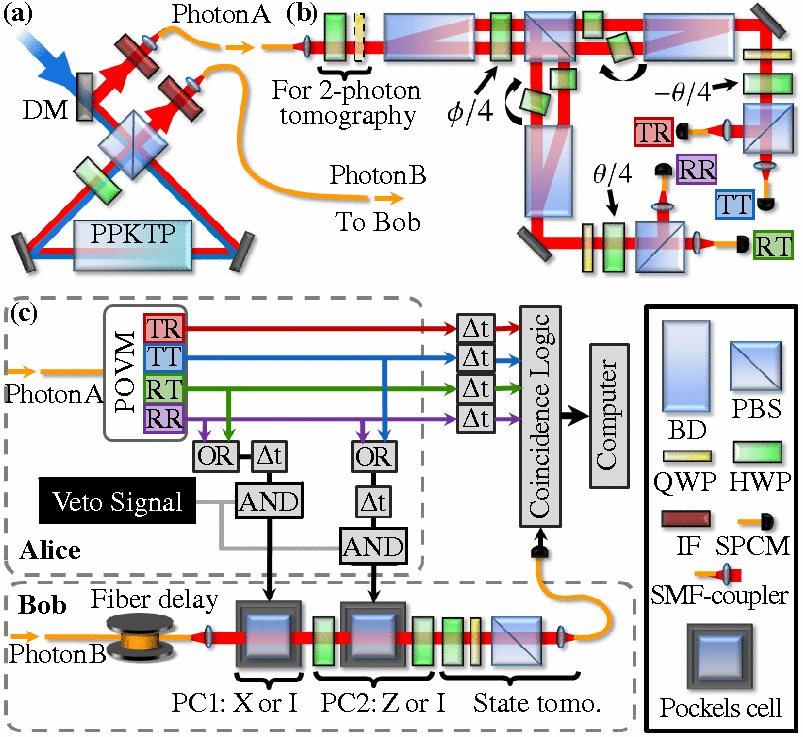}
		\caption{
		\label{fig:experiment}
		(Color online) Remote state preparation experiment.
		(a) Entangled photon pairs are produced via parametric downconversion in a
		polarization-based Sagnac interferometer \cite{kim06a,fedrizzi07a} (see text for details)
		and distributed to Alice and Bob in single-mode fibers (SMFs). (b) The 
		apparatus used by Alice to perform a POVM on her photon. It is 
		a double-interferometer
		based on calcite beam displacers (BDs) which couple the polarization qubit
		to a `path' qubit for the generalized measurement. Futher details are
		in the text. (c) Schematic of the entire experimental RSP protocol. The POVM
		$\{E_m\}$ is performed on Photon A. Based on the outcome $m$ a message
		is encoded in two classical electronic signals, and then sent to Bob with
		probability $r$ which is controled by a Veto Signal. Dependent on the message,
		up to two Pockels cells (PCs) fire to perform the necessary unitary correction
		on Photon B, which has been delayed in a 50~m fiber to allow time
		to trigger the PCs. Bob's output is analyzed using quantum state tomography.
		Note: DM: dichroic mirror; IF: blocking and interference filter; PBS: polarizing
		beamsplitter; H(Q)WP: half (quarter) wave-plate; SPCM: fiber-coupled single-photon
		counting module.
		}
\end{figure}

\subsection{Implementation}\label{ssec:implementation}
Our experimental setup comprises four parts:
\begin{enumerate}
\item The source for preparation of the Bell state $\ket{\Phi^+}_{AB}=\PhiPlusPol$. 
\item The apparatus for performing Alice's POVM $\{E_m(\phi,\theta)\}$ on her qubit and subsequent logic to determine the message sent to Bob. 
\item Bob's implementation (via Pockels cells) of the unitary correction $\sigma_m$ on his qubit. 
\item Tomographic analysis of the resulting remotely prepared qubits.
\end{enumerate}

The source of entangled photon pairs is shown in Fig. \ref{fig:experiment}a) \cite{kim06a, fedrizzi07a}. It relies on spontaneous parametric downconversion in a polarization-based Sagnac interferometer. A 25~mm long periodically-poled KTiOPO$_4$
(PPKTP) nonlinear optical crystal is embedded in the interferometer, and bi-directionally pumped by a grating-stabilized diode laser outputting $<1$~mW
at 404.5 nm. The crystal is temperature-tuned for collinear, degenerate type-II quasi-phase-matching. The resulting entangled photon pairs are coupled into
single-mode fibers (SMFs) which, at this pump power, typically yield singles (coincidence) rates
of  150~(30)~kHz. Polarization controllers (bat ears, not shown) in the fibers
rotate the output to the Bell state $\ket{\Phi^-}=\frac{1}{\sqrt{2}}\left(\ket{HH}-\ket{VV}\right)$. However, the first HWP in Alice's POVM apparatus, when not in use for tomographic characterization of the source output, flips this state to $\ket{\Phi^+}$ before
her measurement.

The apparatus for performing Alice's POVM is shown in Fig. \ref{fig:experiment}b), and
has been described in detail in Ref. \cite{biggerstaff09a}. It is a polarization-based double interferometer employing waveplates, polarizing beamsplitter cubes (PBSs), and calcite beam-displacers. The latter serve to couple polarization with path,
thereby enlarging the state space for the generalized measurement \cite{ralph07a,lanyon09a}; half-wave plates
(HWPs) are used to set the parameters $\{\phi,\theta\}$ of the POVM $\{E_m\}$. The four output modes are coupled into SMF and detected with single-photon counting modules (SPCMs). The outcome TR which stems from (T)ransmission at the first PBS and (R)eflection at the second corresponds to $\frac{1}{2}\rho(\phi,\theta)$, and the RT, RR, and TT outcomes to $\frac{1}{2}Z\rho Z$, $\frac{1}{2}X\rho X$, and $\frac{1}{2}XZ\rho Z X$, respectively.

Due to its the beam-displacer-based construction, the POVM apparatus is inherently phase-stable, without need for active stabilization \cite{obrien03a}. To set the appropriate phase in each interferometer arm, we use an 809~nm diode laser injected through the input SMF and a removable polarizing optic;
following careful tuning we typically measure classical interference visibilities $>99.8\%$. The phase itself is set via tilting HWPs at $0\dg$, shown with arrows in \ref{fig:experiment}b), in one path of each interferometer arm about their vertical axes. 
 This phase need only be set periodically, typically once per day, and can stay set for four hours or more, provided the ambient lab temperature remains stable, allowing time for the preparation and tomography of several hundred different states.

By means of fast electronic logic gates, the POVM outcome is encoded into the voltage state of two TTL signals. The POVM outcomes TR, TT, RT, and RR are encoded as binary strings ``00'', ``01'', ``10'', and ``11'', respectively. A value of ``1'' for the first (second) bit corresponds to a TTL pulse which will trigger Bob's X (Z) correction. For the preparation of pure states, these signals are always transmitted to Bob.

For the preparation of mixed states, a Veto Signal (see Fig. \ref{fig:experiment}c)) is produced with probability $(1-r)$, which is
set by the fraction of time spent in the ``0'' state of a 2~MHz TTL square-wave from a function generator. If a veto signal is generated, it blocks the transmission of TTL pulses to Bob, which is equivalent to sending the message ``00''.
As a result, the message ``00'' is sent with probability $(1-r)+r/4$, and the other three messages are each sent with probability $r/4$. Because the arrival times of photons in Alice's POVM apparatus are random, and the total rate much less than 2~MHz, the decision whether to send each message is both random and independent.

Depending on the message received, Bob must implement the unitary
correction $\sigma_m$ on his qubit in order to achieve deterministic RSP. As shown in Fig. \ref{fig:experiment}c), this is accomplished by means of two fast RbTiOPO$_4$ (RTP) Pockels cells (Leysop RTP4-20-AR800). At 809~nm these have a half-wave voltage of 1.027~kV and a switching time of $<5$~ns. The first Pockels cell is oriented so that when triggered it implements an $X$ operation, which flips $\ket{H}\mapsto\ket{V}$, and is fired if the first bit of the received messages is ``1''. The second Pockels cell, though similarly oriented, is preceeded and followed by HWPs at 22.5\dg which rotate its action to $Z$, so that it flips $\ket{D}=\ket{H}+\ket{V}\mapsto\ket{A}=\ket{H}-\ket{V}$. This Pockels cell is fired if the second bit of the received message is ``1''. When not fired, each Pockels cell and associated waveplates have no net affect on the polarization of transmitted photons (they perform $\iden$.) Bob's photon is stored in a 50~m loop of SMF to allow time for Alice's POVM and logic operations, and triggering of the Pockels cells based on the message.

After the correction stage, Bob's remotely prepared qubit is analyzed
using a polarization analyzer consisting of a QWP, HWP, and PBS, fiber-coupled to a SPCM. Coincidence counts between this detector and Alice's detectors, summed over her four POVM outcomes, yields our raw data. Note that we do not subtract `accidental' coincidence detections. For each remotely prepared state, the measured output density matrix $\rhooutm$ is tomographically reconstructed using a maximum-likelihood technique \cite{james01a} based on coincidence measurements for each of six settings of Bob's analyzer (the eigenstates of $X$, $Y$, and $Z$).

\subsection{Results}\label{ssec:results}

\begin{figure}
\centering
\includegraphics[width=1 \columnwidth]{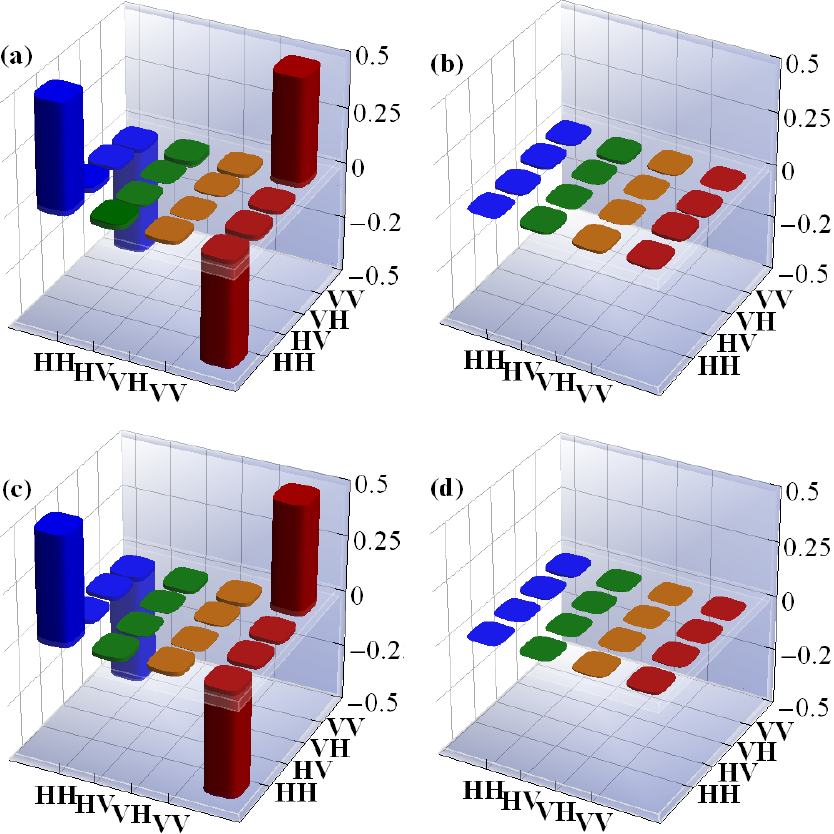}
\caption{
\label{fig:densitymatrices}
(Color online) Experimentally reconstructed density matrices of 
our 2-photon entangled state: real part (left) and imaginary part
(right). The top row \{a) and b)\} represents the source state, as aligned
for the remote preparation of the pure states (Fig. \ref{fig:comparison});
the bottom row \{c) and d)\} represents the source state as aligned (on a subsequent day) for
the preparation of mixed states (states with $r<1$ in Fig. \ref{fig:mixed} and Table
\ref{tab:mixed}.) The state in the top (bottom) row has fidelity $F=0.9807\pm0.0004$ ($0.9813\pm0.0003$) with the ideal state $\ket{\Phi^-}$, tangle $T=0.935\pm0.002$ ($0.932\pm0.001$), and purity $P=\tr(\rho^2) =0.9676\pm0.0009$ ($0.9659\pm0.0007$) \cite{white01a}.
}
\end{figure}

\begin{figure}
	\centering
		\includegraphics[width=1 \columnwidth]{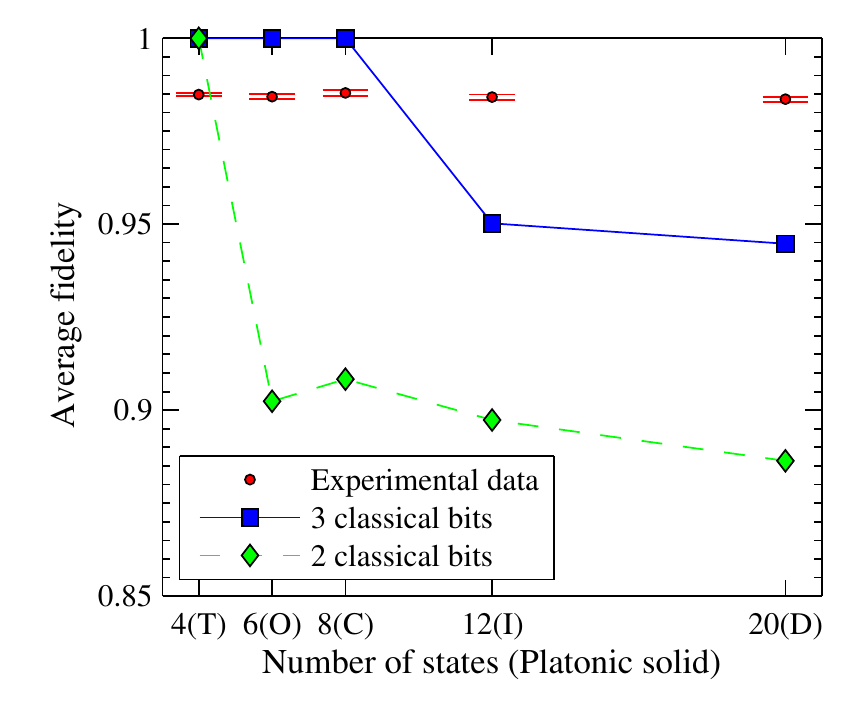}
	\caption{Experimentally-achieved mean fidelities $\<F(\target,\rhooutm)\>$ and optimal classical benchmarks for target ensembles of pure states based on the five Platonic solids shown in Fig. \ref{fig:optimal}. The error bars shown are the standard error of the mean. Any experimental data point above the green diamonds (blue squares) represents results that are not possible with only two (three) cbits communication and no preshared entanglement.  In all cases where the classical benchmark is less than unity, the experimental results surpass the benchmarks conclusively. Note: lines are included only to guide the eye and do not represent calculated thresholds.}
	\label{fig:comparison}
\end{figure}

\begin{figure}
	\centering
		\includegraphics[width=1 \columnwidth]{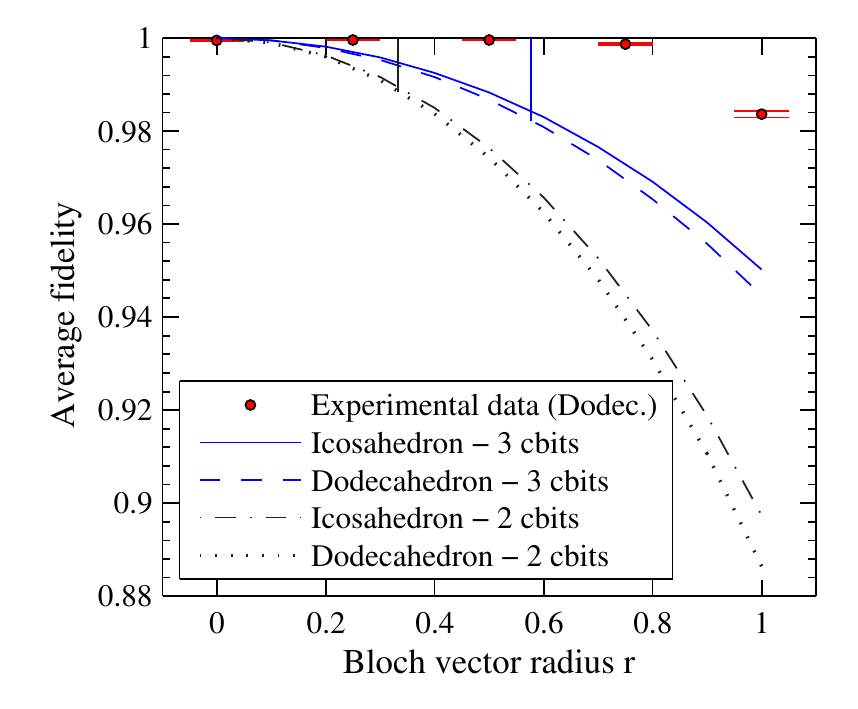}
	\caption{Experimentally-achieved mean fidelities and optimal classical thresholds versus Bloch vector radius. The target ensembles consist of uniform distributions of states of Bloch radius $r$ which form the vertices of either an icosahedron or dodecahedron. The lower (upper) pair of lines are bounds on the classical average fidelity arising from specific two (three) cbit classical strategies. Within each pair, the higher bound is for the icosahedron. Experimental data must lie above and to the right of the bounds to be in the quantum regime, but even points in this region may be possible to achieve without pre-shared entanglement by some non-deterministic classical strategy. Experimental data points for the icosahedron ensembles are similar but were not plotted because at this scale they are not distinguishable from those of the dodecahedron.
		\label{fig:mixed}
}
\end{figure}

Our data was collected over two days: Day 1 for the remote preparation and tomography of all the pure states in Figs. \ref{fig:optimal} and \ref{fig:comparison}, and Day 2 for the mixed states (those where the intended state has $r<1$) in Fig. \ref{fig:mixed} and Table \ref{tab:mixed}.
Each day we characterized the entangled state $\rho_{AB}$ via over-complete state tomography; the results are shown in Fig. \ref{fig:densitymatrices}. The polarization of Photon A is analyzed using the first two waveplates in the POVM apparatus and the $H$-polarized output of the first beam displacer; the other output is blocked and all subsequent waveplates are set so as to direct the resulting photons to the TT POVM outcome. Coincidences are recorded between this output and that of the polarization analyzer following Bob's (switched off) Pockels cells. We perform a tomographically overcomplete set of 36 different measurements, comprising all combinations of the six eigenstates of $X$, $Y$, and $Z$ on Alice's and Bob's qubit, respectively. The results are used to reconstruct the two-photon density matrix \cite{james01a}. On Day 1 (2) 
the measured state $\rho_{AB}$ had fidelity $F=0.9807$ (0.9813) with the ideal output $\ket{\Phi^-}$. When not in use for two-photon tomography, the first QWP on Alice's side is removed, and the HWP is set to $0\dg$, thereby flipping the source output to $\ket{\Phi^+}$ before the POVM.

We remotely prepared and tomographically reconstructed 178 qubit states, and for each measured output density matrix $\rhooutm$ we calculate its agreement with the target state $\target$ using the fidelity (Eq. \ref{eq:fidelity}). Our mean fidelity $\<F(\target,\rhooutm)\>$ is 0.9951, and all but three of the 178 states have $F>0.98$.  However, following Ref. \cite{peters05a}, we also calculate the expected remotely prepared state $\rhoexp$ based on our measured imperfect two-photon entangled state $\rho_{AB}$ as shown in Fig. \ref{fig:densitymatrices}, but assuming perfect operation of the POVM and unitary correction. If we then examine the fidelity $F(\rhoexp,\rhooutm)$, we obtain
$\<F\>=0.9995$, with 177 out of the 178 states having $F>0.9975$. However, in the remainder of the paper and in Figs. \ref{fig:optimal}, \ref{fig:comparison}, \ref{fig:mixed}, and Table \ref{tab:mixed}, the fidelities we report are the more stringent $F(\target,\rhooutm)$, as these are appropriate for comparison with the bounds on classical RSP.

In order to test our experimental RSP implementation against the benchmarks derived in Sec. \ref{sec:thresholds}, we prepare pure states with settings $\{\phi,\theta\}$ corresponding to the vertices of the five Platonic solids inscribed in the Bloch sphere. The orientations used for each polyhedron, along with the results for $\<F\>$, are shown in Fig. \ref{fig:optimal}. These results are compared graphically against the corresponding benchmarks for classical RSP with two and three cbits in Fig. \ref{fig:comparison}. In all instances where the benchmark based on classical RSP is less than unity, our experimentally-determined values surpass it conclusively. This confirms that our results cannot be produced without shared entanglement, even when a comparable amount of classical communication is allowed.

\begin{table*}
 	\begin{tabular}{| c || c | c | c | c | c |}
 		\hline\hline
    $r$ & $0.00$ & $0.25$ & $0.50$ & $0.75$ & $1.00$ \\ 
    \hline\hline 
		Icosahedron & $0.99944 \pm 0.00008$ & $0.99967 \pm 0.00007$ & $0.99964 \pm 0.00008$ & 				$0.9987 \pm 0.0002$ & $0.9841 \pm 0.0007$ \\ 
		\hline 
    Dodecahedron & $0.9995 \pm 0.00006$ & $0.99961 \pm 0.00003$ & $0.99963 \pm 0.00004$ & 				$0.9987 \pm 0.00015$ & $0.9836 \pm 0.0007$ \\
    \hline\hline 
  \end{tabular}
 \caption{Experimentally achieved average RSP fidelities $\<F(\target,\rhooutm)\>$
 versus Bloch vector radius $r$. The data is for ensembles with settings $\{\phi,\theta\}$
 corresponding to the vertices of an icosahedron and dodecahedron with varying outsphere radii $r$.
 The reported uncertainty is the standard error of the mean.
 \label{tab:mixed}
 }
\end{table*}

To test our ability to prepare arbitrary mixed qubit states, we use settings $\{r,\phi,\theta\}$, where $\phi$ and $\theta$ correspond to the vertices of the icosahedron and dodecahedron, for all $r \in \{0.00,0.25,0.50,0.75,1.00\}$. Our results
for $\<F\>$ are summarized in Table \ref{tab:mixed}. In Fig. \ref{fig:mixed} these results are compared to the lower bounds found in Section \ref{sec:thresholds} for classical RSP; again our data surpass the bounds on classical RSP whenever the benchmark is less than unity.

\section{Conclusion}\label{sec:conclusion}

We have investigated the theory of remote state preparation involving only classical communication resources. Based on sets of states forming the vertices of Platonic solids we derived several fidelity-based RSP benchmarks bounding such classical protocols. We have described and implemented an experimental protocol for high-fidelity, fully deterministic remote preparation of arbitrary photonic qubit states, and compared its results with our benchmarks. The results show a clear violation of all the classical thresholds whenever the classical protocols do not trivially allow for a perfect RSP strategy. We also examined the special cases where the states to be remotely prepared are i) chosen uniformly from all pure qubit states and ii) mixed qubit states, all with Bloch radius $r$. We provided appropriate benchmarks in these cases, and our experimental fidelity values once again surpass all sub-unity benchmarks.

An interesting extension to this work would be to deal with the problem of post-selection. Because of device inefficiencies and losses, many more photons are used in the experiment than are actually counted in the final analysis. Carefully counting the lost photons and allowing Alice and Bob to use them as comparable classical resources, without loss, would lead to more difficult benchmarks. Yet it would also provide even stronger support for claims that an experiment evidences genuine quantum behaviour. As well, a comparison of experimental data with the classical thresholds, which were derived for ideal conditions, necessarily has to assume fair sampling \cite{clauser69a,clauser74a}.
\acknowledgments{We thank G. O. Myhr, D. Hamel, and G. Weihs for valuable discussions, and Z. Wang for designing and building our TTL logic. D.B. and R.K. are thankful for support from the Mike and Ophelia Lazaridis Fellowship and IQC, respectively. We are grateful for financial support from NSERC (Discovery Grant and QuantumWorks), OCE, and CFI.} 

\appendix

\section{Proof of bound for continuous case}\label{app:continuumbound}
In this appendix, we proof two claims from section \ref{ssec:continuum} used to find upper bounds on the classical threshold for states on the Bloch sphere. 

\emph{Claim 1.} Define a circular cap as the set of points on the surface of the Bloch sphere lying north of some fixed latitude or any rigid spherical rotation of this. Amongst all partitions of surface area $A>0$, a circular cap has the longest average Bloch vector.

\emph{Proof.} Consider an arbitrary partition of total surface area $A>0$ (we do not assume that this partition is connected). This partition, which we call $\Gamma$, defines some average Bloch vector $\overrightarrow{r_{\Gamma}}$. We will compare partition $\Gamma$ with a circular cap of area $A$ centred along the direction of $\overrightarrow{r_{\Gamma}}$, which we shall denote by $C$. Partition $C$ has an average Bloch vector $\overrightarrow{r_C}$. If $\overrightarrow{r_{\Gamma}}$ is the zero vector, then $r_C\geq r_{\Gamma}=0$. If not, then without loss of generality we can assume $\overrightarrow{r_{\Gamma}}$ points along the $z$-axis. By construction, $\overrightarrow{r_C}$ must also point along the $z$-axis. Using $\Gamma$ and $C$, the Bloch sphere can be divided into four disjoint regions: $R_1 = \Gamma \cap C$, $R_2 = \Gamma - C$, $R_3 = \Gamma^c \cap C$, and $R_4 = \Gamma^c - C.$ Each of these regions has average Bloch vector $\overrightarrow{r_i}$ and area $A_i$, $i=1,...,4$. Also note that we must have $A_2=A_3$. 

Since $\Gamma = R_1\cup R_2$, and $C = R_1\cup R_3$, we must have
\begin{align}
 \overrightarrow{r_\Gamma} & = \frac{A_1}{A_1+A_2}\overrightarrow{r_1} + \frac{A_2}{A_1+A_2}\overrightarrow{r_2},~ \text{and}\\
 \overrightarrow{r_C} & = \frac{A_1}{A_1+A_2}\overrightarrow{r_1} + \frac{A_3}{A_1+A_3}\overrightarrow{r_3}. &
\end{align}
From this we conclude that 
\begin{equation}\label{parallel}
 \overrightarrow{r_C} = \overrightarrow{r_\Gamma} + \frac{A_2}{A_1+A_2}(\overrightarrow{r_3}-\overrightarrow{r_2}),
\end{equation}
i.e. that the vector $\overrightarrow{r_3}-\overrightarrow{r_2}$ also lies along the $z$-axis. However, it might point in the negative $z$-direction.

But the boundary of $C$ lies at some fixed height $H_C$ on the $z$-axis. By construction, every state in $R_3$ has a $z$-component higher than $H_C$ and every state in $R_2$ has a $z$-component lower than $H_C$. Then the $z$-component of $\overrightarrow{r_3}$ must be larger than that of $\overrightarrow{r_2}$. Therefore, their difference $\overrightarrow{r_3}-\overrightarrow{r_2}$ has a positive $z$-component. From Eq. (\ref{parallel}), we can conclude that $\overrightarrow{r_C}$ is longer than $\overrightarrow{r_\Gamma}$. Thus, for fixed area $A$, a circular cap gives the longest average Bloch vector.

\emph{Claim 2.} Knowing that a circular cap gives the optimum Bloch vector length for fixed area $A$, we want to optimize the objective function
\begin{equation}\label{objective}
 \sum_{k=0}^{2^c-1}\langle F \rangle^{\text{max}}_{A_k} = \frac{1}{2}\left( 1 + \sum_{k=0}^{2^c-1}p_k r_k^{\text{max}} \right)
\end{equation}
subject to the constraint
\begin{equation}\label{constraint}
 \sum_{k=0}^{2^c-1} A_k = 4\pi.
\end{equation}
We claim that this is optimized when all areas are equal.

\emph{Proof.} To obtain the optimal Bloch vector as a function of area, we temporarily centre a spherical cap on the $z$-axis and integrate up to some final angle $\phi^f_k$,
\begin{align}
 r^{\text{max}}(\phi^f_k) = &\frac{1}{A_k}\left| \int_{\theta=0}^{2\pi}\int_{\phi=0}^{\phi^f_k} \sin(\phi)\cos(\phi) d\theta d\phi \right|\notag\\ =& \frac{1}{A_k}\pi\sin^2(\phi^f_k).
\end{align}
Reparameterizing using $A_k = 4\pi\sin^2(\frac{\phi^f_k}{2})$, we end up with 
\begin{equation}
 r^{\text{max}}(A_k) = 1-\frac{A_k}{4\pi}.
\end{equation}
The corresponding probabilities are given by $p_k = \frac{A_k}{4\pi}$. We can group the objective function (\ref{objective}) together with the constraint (\ref{constraint}) into the following Lagrange function:
\begin{align}
 \Lambda(A_k,\lambda) = &\frac{1}{2}\left(1+\sum_{k=0}^{2^c-1}\frac{A_k}{4\pi}\left(1-\frac{A_k}{4\pi}\right)\right)\notag\\ + &\lambda\left( \sum_{k=0}^{2^c-1}\frac{A_k}{4\pi} - 1 \right).
\end{align}
Solving this Lagrange problem for the maximum yields $A_k = \frac{4\pi}{2^c}$ for every $k$. Hence, the optimal distribution of areas occurs when all they are all equal.

\section{Proof of optimal average fidelity for mixed states}\label{app:mixedstates}
In this appendix, we prove the optimality of Eq. (\ref{Fmax_mixed}). Since we are dealing with qubits, we can make use of an alternative formula for fidelity found in \cite{miszczak09a}, namely
\begin{equation}
 F(\sigma,\tau) = \tr(\sigma\tau)+\sqrt{1-\tr(\sigma^2)}\sqrt{1-\tr(\tau^2)}.
\end{equation}
Under the assumptions that the target states all have the same Bloch vector length $r$ and that the message strategy is deterministic, the average fidelity is
\begin{align}
 \langle F \rangle 	= & \displaystyle\sum_{\alpha} p_\alpha F(\rho_\alpha^{\text{tar}},\rho^{\text{out}}_{m(\alpha)})\notag\\
			= & \displaystyle\sum_{k=0}^{2^c-1}\displaystyle\sum_{\alpha\in k} p_\alpha F(\rho_\alpha^{\text{tar}}, \rho_k^{\text{out}})\notag\\
			= & \displaystyle\sum_{k=0}^{2^c-1} p_k \displaystyle[\tr(\overline{\rho_k} \rho_k^{\text{out}})\notag\\
			  & + \sqrt{\frac{1-r^2}{2}}\sqrt{1-\tr((\rho_k^{\text{out}})^2)} \displaystyle].
\end{align}
As before, $p_k=\sum_{\alpha\in k}p_\alpha$ is the probability of sending message $k$ and $\overline{\rho_k}=\frac{1}{p_k}\sum_{\alpha\in k}p_\alpha\rho_\alpha$ is the weighted average of states where message $k$ is sent. The quantity in square brackets will be denoted by
\begin{equation}
 G_k[\rho_k^{\text{out}}] = \tr(\overline{\rho_k} \rho_k^{\text{out}}) + \sqrt{\frac{1-r^2}{2}}\sqrt{1-\tr((\rho_k^{\text{out}})^2)}.
\end{equation}

For each $k$, we need to find the choice of $\rho_k^{\text{out}}$ which optimizes $G_k$. Working in the eigenbasis of $\overline{\rho_k}$, we have
\begin{equation}
 \rho_k^{\text{out}} = \begin{bmatrix}
                 a & b\\
                 b* & d
                \end{bmatrix},
\end{equation}
with $a,d\in\mathbb{R},b\in\mathbb{C}$. From the above expression for the fidelity, the optimal choice of $\rho_k^{\text{out}}$ should be simultaneously diagonal with $\overline{\rho_k}$, i.e. $b=0$. Equivalently, the Bloch vectors of $\overline{\rho_k}$ and the optimal $\rho_k^{\text{out}}$ should be parallel. Denoting the magnitudes of these Bloch vectors by $r_k$ and $s_k$, respectively, we are left with
\begin{align}
 G_k = & \frac{1}{4}\left[ (1+r_k)(1+s_k) + (1-r_k)(1-s_k)\right]\notag\\
       & + \sqrt{\frac{1-r^2}{2}}\sqrt{\frac{1-s_k^2}{2}}\notag\\
     = & \frac{1}{2}\left( 1+r_k s_k \right) + \sqrt{\frac{1-r^2}{2}}\sqrt{\frac{1-s_k^2}{2}}.
\end{align}
Since $r_k$ is fixed by the choice of target state partitioning, we differentiate $G_k$ with respect to $s_k$ and find where this derivative equals zero. The result is 
\begin{equation}
 s_k = \pm \frac{r_k}{\sqrt{r_k^2+1-r^2}}.
\end{equation}
The positive root will give the maximum of $G_k$, which works out to be
\begin{equation}
 G_k^{\text{max}} = \frac{1}{2}\left( 1+\sqrt{r_k^2+1-r^2} \right).
\end{equation}
Collecting all the terms together yields Eq. (\ref{Fmax_mixed}).

\bibliography{rsp}

\end{document}